# Linguistic Analysis of Requirements of a Space Project and their Conformity with the Recommendations Proposed by a Controlled Natural Language

Anne Condamines and Maxime Warnier

CLLE-ERSS, CNRS and Université Toulouse 2 – Le Mirail / CNES
{anne.condamines,maxime.warnier}@univ-tlse2.fr

**Abstract.** The long term aim of the project carried out by the French National Space Agency (CNES) is to design a writing guide based on the real and regular writing of requirements. As a first step in the project, this paper proposes a linguistic analysis of requirements written in French by CNES engineers. The aim is to determine to what extent they conform to two rules laid down in INCOSE, a recent guide for writing requirements. Although CNES engineers are not obliged to follow any Controlled Natural Language in their writing of requirements, we believe that language regularities are likely to emerge from this task, mainly due to the writers' experience. The issue is approached using natural language processing tools to identify sentences that do not comply with INCOSE rules. We further review these sentences to understand why the recommendations cannot (or should not) always be applied when specifying large-scale projects.

## 1 Introduction

The study presented in this paper was conducted with a view to improving the writing of requirements at CNES (Centre National d'Études Spatiales).

The CNES and our laboratory have been collaborating for several years on questions concerning terminology, text management and the study of risks related to the use of language [1]. As linguists, we propose methods and results based on a corpus linguistics approach, assisted by tools such as parsers, statistical tools, terminology extractors, concordancers or scripting languages. More recently, we were approached on the specific problem of writing requirements.

The CNES is the French space agency and, as such, is responsible for designing space systems. Therefore, it has to draft specifications (that must clearly and precisely describe its needs) which are intended for companies that respond to the bids; and, in turn, it also responds to bids from other scientific, commercial or military partners. The Quality Department of the CNES, however, is aware that these specifications are not always clear, and that as a result there may be divergent interpretations, leading to additional costs, delays or even litigation (since requirements are part of the contract clauses).

In order to improve the quality of requirements, many projects have been developed by computational researchers to check the consistency of the requirements after they were written (see [2–4], among others). Still, we believe that the writing itself can be improved by proposing a guide closer to the actual way in which engineers write requirements.

In the present study, two kinds of documents were used: the Guide for Writing Requirements recommendations proposed by INCOSE (International Council on Systems Engineering) [5] (a controlled natural language, see below); and a subset of the specifications of a project: Pleiades (see below).

We propose a linguistic diagnosis of the way requirements are written in the project by comparing these requirements with the recommendations of the INCOSE guide.

The point of view underlying our approach is that guides for writing specifications are not fully adapted to the real writing process: they are sometimes too constraining, and sometimes insufficiently so. They are not written by linguists but by domain experts with a prescriptive point of view based on their experience. This is the case for example in the field of air-traffic control where the ICAO (International Civil Aviation Organization) phraseology is written by controllers [6]. Even if these guides are not always adapted to the reality of language use, we consider that they constitute a good starting point because of the experience of the domain experts. Our other starting point is constituted by specifications that are not written following the recommendations of a guide: this is the case at CNES.

Indeed, CNES engineers do not use a controlled natural language in order to write better specifications, only requirement management tools (such as IBM Rational DOORS). Nevertheless, they are all experienced in this type of writing. Thus, even if the writers do not consciously follow a controlled natural language, we assume the existence of regularities in the way they write requirements. Writers are indeed influenced both by existing specifications and by certain spontaneous regularities which tend to occur in each recurrent writing situation, two characteristics attributed to textual genres. According to Bhatia [7], a *textual genre* may be defined as "a recognizable communicative event characterized by a set of communicative purpose(s) identified and mutually understood by the members of the professional or academic community in which it regularly occurs".

It can be noted that the notion of textual genre is not always properly distinguished from that of *sublanguage*. See for instance the definition given by Somers: "A sublanguage is an identifiable genre or text-type in a given subject field, with a relatively or even absolutely closed set of syntactic structures and vocabulary" [8]. Other authors such as Kurzon [9], Temnikova [10] or Kuhn [11] have highlighted this point. Historically, the most important difference is that the notion of sublanguage was proposed by Harris from a mathematical and distributional perspective [12], while that of textual genre comes from a more sociolinguistic approach [7, 13] or even a corpus linguistic one [14]. In both cases, one of the most important characteristics is that linguistic regularities are associated with speakers of the same community. This feature of spontaneous linguistic regularities has been characterized as *normaison* ("norming') by the French Linguistic School of Rouen [15] as opposed to *normalisation* ("normalization") that concerns the case where linguistic norms are imposed by an organism. In short, we could say that our aim is to propose a normalisation based on the identification of normaison, or, in other words, to improve the writing of specifications without imposing a standard that is too far removed from the engineers' natural practice.

The paper comprises two main parts. In the first one (see section 2), we present the tool-assisted method used for making the diagnosis. In the second one (see section 3), we describe and discuss our preliminary results.

## 2  Methodology

Several guides for writing requirements exist, and most of them were designed to avoid undesirable properties of natural language ("unrestricted natural language brings with it a host of well-known problems" [16]), such as ambiguity, polysemy, vagueness, and so on [1, 17].

To ensure that these guidelines are close enough to actual practices, and thus really usable, we decided to carry out a diagnosis of the way the specifications are drafted at CNES and then to compare this process with the recommendations made by one of those guides. The aim is to evaluate the conformity of the requirements to the recommendations, and see if the latter can be brought closer to reality.

We will first briefly describe our corpus of requirements and the tools we used, and then the linguistic phenomena selected for study in the controlled natural language that we used as a reference.

### 2.1  Description of the corpus

A subset of the specifications of an Earth observation satellite called Pleiades, launched in 2011, was obtained from the CNES. From these specifications, we extracted the requirements, that is to say only those parts that play the role of contractual obligations between the CNES and its subcontractors. Requirements should not contain unnecessary information, such as examples or comments.

Requirements are intended to be autonomous; they are therefore supposed to have no link with the textual segments which precede or follow them. In the specifications we were given, the requirements were easily identifiable because they were framed by specific tags.

The requirements were all written in natural language, but some also contained tables or diagrams (which were removed, since they cannot be analyzed automatically). In theory, they should be fully understandable even without those figures – but in practice, this is not always the case.

The resulting corpus is composed of 1,142 requirements (nearly 53,000 words) in French.

## 2.2 Tools and resources

We used several tools to perform the tasks described in section 3. The syntactic analysis was done using Talismane [18], an open-source parser developed in our laboratory, while the open-source corpus processor Unitex [19] was used for sentence chunking. Short handmade Perl scripts were written for other needs (extraction of the requirements, detection of long sentences, and so on).

We also compared our corpus to two other corpora (reduced to the exact same size): (1) a handbook written by experts from the CNES about techniques and technologies used for building and operating spacecraft, intended for semi-experts, and (2) some articles from the French national newspaper *Le Monde*.

## 2.3 INCOSE recommendations

In order to compare the requirements corpus with a controlled natural language, we used the Guide for Writing Requirements recommendations proposed by INCOSE. The aim of this guide is presented as follows: "to draw together advice from a variety of existing standards into a single, comprehensive set of rules and objectives" (p. 10). It is quite general since it "is intended to cover the expression of requirements from across disciplines" (p. 12). INCOSE is therefore intended for engineers who write or review requirements. It can be clearly considered a "naturalist" controlled language (as opposed to the "formalist" approach) [20], whose goal is to facilitate human-to-human communication [21].

Like many other controlled natural languages (CNL) aimed at improving communication among humans, the main purpose of INCOSE is to ensure that the message written in natural language has only one possible interpretation. It is worth noting that this point of view about natural language is far from the one adopted by linguistics.[1] It can be reasonably assumed, however, that by establishing guidelines in narrowly-defined situations, it may be possible to limit (if not to remove completely) the inherent difficulties linked to natural language such as ambiguity.

INCOSE has the four characteristics of controlled natural languages proposed by Kuhn [11], since it has one base language (English), it is a constructed language, it sets constraints on the vocabulary, the syntax and the semantics, and the resulting textual requirements are still understandable by English speakers.

It is not a mere style guide, because the recommendations are real rules, not hints – even if the authors admit that "rules have to constantly be adapted to particular situations". All of them are followed by objectives that explain why the rules are useful.

The main "objectives for writing requirement statements" are: singularity, completeness, necessity, comprehensibility, concision, precision and non-ambiguity.

---

[1] According to Jakobson, for example, the referential function, which is the closest to the one consisting in transmitting information, is only one among the six functions of language [22].

These recommendations are translated into linguistic instructions. We selected several of these instructions and analyzed our corpus to see how often they appear.

Because the phenomena we chose to observe are quite general (i.e. not highly language-dependent), we assume that most of the conclusions we propose for French are valid for English as well. In fact, INCOSE, while written in English and mainly based on older English guides, sometimes gives examples in French.

Since it was not possible to check the conformity of the requirements to all the recommendations proposed by INCOSE (partly because the study is still in its initial stage, and partly because several of the recommendations cannot be verified in an automated manner), we decided to focus on a selection, all related to what could be called "comprehensibility"; that is, the fact that every (sentence composing a) requirement should be easily understandable by the reader, and that it cannot be misinterpreted, i.e. given a different meaning from the one originally intended by the writer. This notion is closely connected to that of complexity: the more complex a sentence is, the less easy it is to understand.

The first rule from INCOSE that we chose to examine is called "Singularity/Propositionals" and states that "combinators" must be avoided: *"Combinators are words that join clauses together, such as 'and', 'or', 'then', 'unless'. Their presence in a requirement usually indicates that multiple requirements should be written."* Nevertheless, some of them are still present in the examples of "acceptable" specifications; this paradox suggests that the "combinators" cannot always be avoided.

The second rule is called "Completeness/Pronouns" and states that it is better to repeat nouns in full, rather than using pronouns to refer to nouns in other statements: *"Pronouns are words such as 'it', 'this', 'that', 'he', 'she', 'they', 'them'. When writing stories, they are a useful device for avoiding the repetition of words; but when writing requirements, pronouns should be avoided, and the proper nouns repeated where necessary."* However, there is no indication about the conditions required for this repetition to be "necessary"; we can merely infer that the aim is to avoid problems due to anaphora resolution. Besides, in the only example given by INCOSE[2], the ambiguity lies in a determiner, not in a pronoun.

We can already point out that these two rules are very general and seem way too restrictive, and that their justifications are evasive.

## 3  First results

In subsection 3.1, we present our results concerning the frequency of conjunctions, pronouns and long sentences in our corpus. In subsection 3.2, we propose a selection of examples that break the two rules from INCOSE and try to classify them according to their necessity (mandatory, useful or undesirable).

---

[2]  "The controller shall send the driver's itinary (sic) for the day to the driver" must be preferred to "The controller shall send the driver his itinary (sic) for the day".

### 3.1 Quantitative analysis

Thanks to the syntactic analysis, we were able to retrieve all the occurrences of the so-called combinators (since no exhaustive list was given, we looked for all coordinating and subordinating conjunctions) and all the pronouns in the corpus. As can be seen from table 1, both are numerous, suggesting that they are common in unrestricted natural language.

Still, they are much less frequent in requirements than in the other two corpora, handbooks and newspapers. This is particularly clear in the case of pronouns, which are nearly three times more frequent in newspapers (where repetition is seen as an error of style in French) than in requirements (which are usually much shorter). We believe that such a marked difference is an argument in favor of our initial hypothesis that regularities spontaneously arise in daily practice, and that requirement writing can be considered a textual genre, even when not taught as such.

**Table 1.** Number of conjunctions and pronouns in the three corpora

|  | Coordinators | Conjunctions Subordinators | (total) | Pronouns |
|---|---|---|---|---|
| Requirements | 882 (1.66%[3]) | 365 (0.69%) | 1247 (2.35%) | 986 (1.86%) |
| Handbook | 1455 (2.75%) | 442 (0.83%) | 1897 (3.58%) | 1554 (2.93%) |
| Newspaper | 1274 (2.40%) | 579 (1.09%) | 1853 (3.50%) | 2710 (5.11%) |

Finally, we also considered the length of the sentences composing the requirements. Although INCOSE simply recommends "concise" requirements, several guides for technical writing (such as ASD Simplified Technical English [23]) impose a word limit for each sentence4, because it is believed that longer sentences are harder to process. The results of our measures are shown in table 2.

**Table 2.** Length of sentences in the three corpora

|  | # sentences | # sentences with > 25 words | Average sentence length (# words) |
|---|---|---|---|
| Requirements | 4859 | 350 (7.2%) | 11 |
| Handbook | 3456 | 591 (17.1%) | 15 |
| Newspaper | 2201 | 839 (38.1%) | 24 |

Once again, significant differences exist between the three types of documents: sentences tend to be shorter in requirements, and much longer in newspaper articles.

---

[3] Percentages indicate the number of occurrences in relation to the total number of words.
[4] Usually around 20 words for English. We arbitrarily decided that long sentences (in French) are composed of more than 25 words, and that a new sentence begins after each line break.

However, long sentences are not rare in the requirements corpus; there is even one unusually long sentence containing over 70 words:

*"Si la différence (en valeur absolue) entre les dates de fin de lecture de deux fichiers, lus sur tranche de COME M - canal TMI i et sur tranche de COME N - canal TMI j, est inférieure à OPS_DELAI_INTER_FIN_LEC secondes, alors il est interdit d'enchaîner (lecture enchaînée) par la lecture de la tranche de COME N sur le canal i et de la tranche de COME M sur le canal j."*

### 3.2  Qualitative analysis (analysis of examples)

As a first step in the diagnosis, we focus on the description of some examples of sentences that do not follow the INCOSE recommendations and try to understand why.

**Combinators**

*Some combinators are mandatory:*
(1) *"Le générateur de TCH vérifiera **que** la valeur du champ PHASE est comprise entre 0 **et** FREQ_DIV -1."* [*"The generator of TCH will check **that** the value of the field PHASE is between 0 **and** FREQ_DIV-1"*]

In example 1, the subordinating conjunction "que" cannot be avoided, since it introduces the dependent clause[5], and the coordinating conjunction "et" is necessary to set the lower and higher limits of the interval.

*Some combinators are not mandatory, but prevent repetitions and multiple sentences:*
(2) *"Les champs SM_ID **et** FM_ID seront extraits à partir de la BDS"* [*"Fields SM_ID **and** FM_ID will be extracted from the BDS"*]

If the use of "et" were not allowed in example 2, two distinct sentences would be necessary ("Le champ SM_ID sera extrait à partir de la BDS." and "Le champ FM_ID sera extrait à partir de la BDS."). This would lead to longer and probably more confusing requirements: since the two sentences differ by only a single character, the reader may not notice the difference and think it is a duplicated sentence.

However, longer sentences may become less readable:
(3) *"Cette TC permet de passer contrôle thermique plate-forme en mode REDUCED, c'est-à-dire de sélectionner des seuils de régulation "larges" pour le contrôle thermique grossier (pour limiter la puissance consommée), **et** de modifier la valeur d'écrêtage de la puissance injectée pour le contrôle thermique fin."* [*"This TC makes it possible to switch the heat control of the platform to REDUCED mode, i.e. to select "broad" regulation thresholds for a coarse heat control (to limit the power consumed), **and** to change the cut-off value of the injected power for precise heat control."*]

In example 3, it would have been better to clearly distinguish the two actions permitted by the TC – for example, with a bullet list.

---

[5]  In French, the complementizer 'que' must always be used.

*Some combinators provide logical information that may help the reader to better understand the requirements:*

(4) *"pour n=2 la loi de la taille est respectée de fait **mais** le test 'FIFO vide' reste nécessaire"* [*"for n=2 the size rule is always respected, **but** the 'empty FIFO' test is still required"*]

In example 4, the reader is certain that the test is necessary in all cases. Without the first main clause and the logical connector "mais", he could have doubted it.

Nonetheless, in several cases, the use of a coordinator does not seem justified; in particular when two sentences are coordinated by "and":

(5) *"Le format des données de mesure angulaire et Doppler est conforme au standard CCSDS décrit dans le document DA9 **et** le schéma XML respecte le standard décrit dans DA11."* [*"The data format of the angular and Doppler measurement is in accordance with the CCSDS standard described in document DA9 **and** the XML schema complies with the standard described in DA11."*]

(6) *"Les demandes sont saisies sur le FOS **et** le logiciel ARPE gère les conflits entre les demandes Spot, Hélios et Pléïades."* [*"The requests are to be entered on the FOS **and** the ARPE software manages conflicts between the requests from Spot, Hélios and Pléïades"*]

In examples 5 and 6, there is no apparent reason why separate sentences should not be used (parataxis).

*In some cases, problems arise because of the (absence of proper) coordinators:*

(7) *"Pour cela, on utilisera les données BDS (LENGTH et LOCATION_UNIT) de la table des OBCD (globaux) **ou** la description (LONGUEUR) des paramètres diagnostic déjà crées."* [*"For this, we will use the BDS data (LENGTH and LOCATION_UNIT) from the (global) OBCD table **or** the description (LONGUEUR) of the already created diagnostic parameters"*]

In example 7 above, there are two possible solutions (alternative), but no explanation is given to the reader to tell him in which case(s) one of them should be preferred (or whether they are in fact identical).

(8) *"Sur réception de cette TC, le LVC met à jour la table des surveillances standards de l'application destinataire **et** ré-initialise le compteur d'erreur (remise à 0) associé à cette surveillance."* [*"Upon reception of this TC, the LVC updates the table of standard surveillances of the destination application **and** resets the error counter associated to this surveillance"*]

In example 8, we know that the LVC has to do two distinct operations, but it is not clear whether they are supposed to be done at the same time or one after the other.

(9) *"(eg : 2 **et** 10 **ou** 3 **et** 11)"* [*"e.g. : 2 **and** 10 **or** 3 **and** 11)"*]

In example 9, the priorities of the logical operators "et" and "ou" are not clear.

(10) *"Cet ordre est rejeté **si** :* [*"This order is rejected **if**:"*]

    *- le mode NORM automatique est actif*
    *- le satellite est en mode MAN*
    *- le satellite n'est pas en mode convergé (GAP ou SUP)*
    *- un ordre MAN/CAP est déjà en attente d'exécution"*

In example 10, the absence of coordinators between the items in the list is the source of uncertainty: is the order rejected if any of the following conditions is met ("or"), or only if they are all met ("and")? Lists of this kind are very common in our corpus.

**Pronouns**

*Some pronouns must be avoided, because otherwise the requirement is no longer autonomous:*
> (11) *"**Il** calculera aussi, a une fréquence paramétrable (ordre de grandeur 1 mois), la moyenne de mise en œuvre et la comparera à la moyenne maximum afin d'anticiper un problème éventuel."* [*"**It** will also calculate, at a frequency that can be parameterized (at monthly intervals), the average time for commissioning and will compare it to the maximum average in order to anticipate any problems."*]

The requirement given in example 11 cannot be understood by itself, because the pronoun "il" ("it") refers to the subject defined in the previous requirement. (And in another requirement, a reference is made to a "previously stated rule", but there is no indication as to which rule is meant.)

*Some pronouns are mandatory:*
> (12) *"Sur réception de cette TC, le LVC met à jour le paramètre **qui** donne la taille maximum d'un paquet TM de type dump"* [*"Upon reception of this TC, the LCV updates the parameter **that** gives the maximum size of a TM dump packet"*]

Without the relative pronoun "that", it would not be possible to specify which parameter is referred to in example 12.
> (13) *"Il ne sera pas utile de vérifier ce paquet " vide ""* [*"**It** won't be necessary to check that "empty" packet"*]

Impersonal pronouns like the one given in example 13 are widespread in our corpus and can hardly be avoided. They do not refer to another noun.

*Some pronouns are not mandatory, but prevent unnecessary repetitions of words:*
> (14) *"La liste des TCD est définie en BDS. **Elle** est donnée ici à titre informatif:"* [*"The list of TCD is defined in BDS. **It** is given here for information:"*]

Compare example 14 with the same sentences without a pronoun: "La liste des TCD est définie en BDS. La liste des TCD est donnée ici à titre informatif:". [*"The list of TCD is defined in BDS. The list of TCD is given here for information:"*]
> (15) *"Le paquet ne sera généré que s'**il** est activé par le LVC."* [*"The packet will be generated only if **it** is activated by the LVC"*]

Example 15 seems even less natural if rewritten without a pronoun: "Le paquet ne sera généré que si le paquet est activé par le LVC." [*"The packet will be generated only if the packet is activated by the LVC"*]

Moreover, French demonstrative pronouns make it possible to avoid ambiguity between the subject and the object of a sentence:

(16) *"Le générateur de TC ne rejettera pas la création du PARAM_ID diagnostic si **celui-ci** est déjà défini à bord."* [*"The TC generator will not reject the creation of the PARAM_ID diagnostic if **the latter** is already defined on board"*]

In example 16, "celui-ci" refers to the closest noun and is therefore unambiguous, whereas "il" could have been ambiguous.

## 4      Conclusions and future work

We analyzed a corpus composed of genuine requirements that had been written and used by engineers of the CNES to design a space system. We showed that, even if they did not explicitly follow guidelines, their texts have some interesting particularities, such as shorter sentences than in other textual genres.

We also examined two rules (concerning conjunctions and pronouns) from INCOSE, a guide for writing requirements. Using several examples from our corpus, we considered cases where those rules were justified and others where they were inapplicable (at least if literally applied) and should be refined. In fact, we believe that INCOSE, like the guides it is based on, lacks proper linguistic foundations and is not close enough to engineers' real practices. For instance, the recommended absence of pronouns from the requirements – which implies sometimes cumbersome repetitions – seems hardly compatible with its ideal of "concision" (itself seen as "an aid to Comprehensibility, and therefore subsumed by it", p. 16).

In the future, we intend to conduct a deeper linguistic analysis of our results and to focus on terminology so as to study the use and evolution of terms between comparable corpora. We also want to test the rules of INCOSE on another corpus of requirements. More generally, our intention is to inventory all existing rules in French CNL and to try to automatically test them with our corpora. The final step will be to propose a set of rules that is more consistent and closer to established practice in requirements writing.

**Acknowledgments.** We would like to thank the CNES for their active cooperation as well as for providing us with the requirements corpus. We are also grateful to the four reviewers for their relevant suggestions and references.